\documentclass{article}

\usepackage{amssymb,amsmath}
\usepackage{graphicx}
\usepackage{wrapfig}
\usepackage{verbatim}
\usepackage{pstricks}

\def \1 { {\mathbf{1}} }

\newcommand{\pd}[2]{\frac{\partial #1}{\partial #2}}

\newcommand{\mb}{\mathbf}

\def \1 { {\mathbf{1}} }

\title{
Hamiltonian Monte Carlo with Reduced Momentum Flips
}
\author{Jascha Sohl-Dickstein\\
Redwood Center for Theoretical Neuroscience\\
University of California at Berkeley
}
\begin{document}

\maketitle

Hamiltonian dynamics with partial momentum refreshment, in the style of \cite{Horowitz1991}, explore the state space more slowly than they otherwise would due to the momentum reversals which occur on proposal rejection.  These cause trajectories to double back on themselves, leading to random walk behavior on timescales longer than the typical rejection time, and leading to slower mixing.  I present a technique by which the number of momentum reversals can be reduced.  This is accomplished by maintaining the net exchange of probability between states with opposite momenta, but reducing the rate of exchange in both directions such that it is 0 in one direction.  An experiment illustrates these reduced momentum flips accelerating mixing for a particular distribution.

\section{Formalism}\label{formalism} 

A state $\mb \zeta\in R^{N\times 2}$ consists of a position $\mb x\in\mathcal R^N$ and an auxiliary momentum $\mb v\in\mathcal R^N$, $\mb \zeta = \left\{ \mb x, \mb v \right\}$.  The state space has an associated Hamiltonian
\begin{align}
H\left( \zeta \right) &= E\left( \mb x \right) + \frac{1}{2} \mb v^T \mb v
,
\end{align}
and a joint probability distribution
\begin{align}
p\left( \mb x, \mb v \right) &= p\left( \zeta \right) = \frac{1}{Z} \exp\left( -H\left( \zeta \right) \right)
,
\end{align}
where the normalization constant $Z$ is the partition function.

The momentum flip operator $F: \mathcal R^{N\times 2} \rightarrow \mathcal R^{N\times 2}$ negates the momentum.  It has the properties:
\begin{itemize}
  \item $F$ negates the momentum, $F \zeta = F\left\{ \mb x, \mb v\right\} = \left\{ \mb x, -\mb v\right\}$
  \item $F$ is its own inverse, $F^{-1} = F$, $FF \zeta = \zeta$.
  \item $F$ is volume preserving, $\det \left(\pd{ \left( F\zeta\right) }{\zeta^T}  \right) = 1$
  \item $F$ doesn't change the probability of a state, $p\left(\zeta\right) = p\left(F\zeta\right)$
\end{itemize}

The leapfrog integrator $L\left( n, \epsilon \right): \mathcal R^{N\times 2} \rightarrow \mathcal R^{N\times 2}$ integrates Hamiltonian dynamics for the Hamiltonian $H\left( \zeta \right)$, using leapfrog integration, for $n\in\mathcal Z^+$ integration steps with stepsize $\epsilon\in\mathcal R^+$.  We assume that $n$ and $\epsilon$ are constants, and write this operator simply as $L$.  The leapfrog integrator $L$ has the following relevant properties:
\begin{itemize}
  \item $L$ is volume preserving, $\det \left(\pd{ \left( L\zeta\right) }{\zeta^T}  \right) = 1$
  \item $L$ is exactly reversible using momentum flips, $L^{-1} = FLF$, $\zeta = FLFL\zeta$
\end{itemize}

During sampling, state updates are performed using a transition operator $T\left( r\right): \mathcal R^{N\times 2} \rightarrow \mathcal R^{N\times 2}$, where $r\sim U\left( [0, 1) \right)$ is drawn from the uniform distribution between 0 and 1,
\begin{align}
T\left( r\right) \zeta
 =
	\left\{\begin{array}{ccc}
L \zeta & & r < P_{leap}\left( \zeta \right) \\
F \zeta & & P_{leap} \leq r < P_{leap}\left( \zeta \right) + P_{flip}\left( \zeta \right) \\
\zeta    & & P_{leap} + P_{flip}\left( \zeta \right) \leq r
	\end{array}\right.
\label{trans}
.
\end{align}
$T\left( r\right)$ additionally depends on an acceptance probability for the leapfrog dynamics, $P_{leap}\left( \zeta \right)\in[0,1]$, and a probability of negating the momentum, $P_{flip}\left( \zeta \right)\in[0,1-P_{leap}\left( \zeta \right)]$.  These must be chosen to guarantee that $p\left( \zeta \right)$ is a fixed point of $T$.\footnote{This fixed point requirement can be written as $p\left( \zeta \right) = \int d{\zeta'} p\left( \zeta' \right) \int_0^1 dr \delta\left( \zeta - T\left(r\right) \zeta' \right)$.}

\begin{figure}
\begin{center}
\parbox[c]{0.95\linewidth}{
\begin{tabular}{cc}
\begin{tabular}{c}\includegraphics[width=0.46\linewidth]{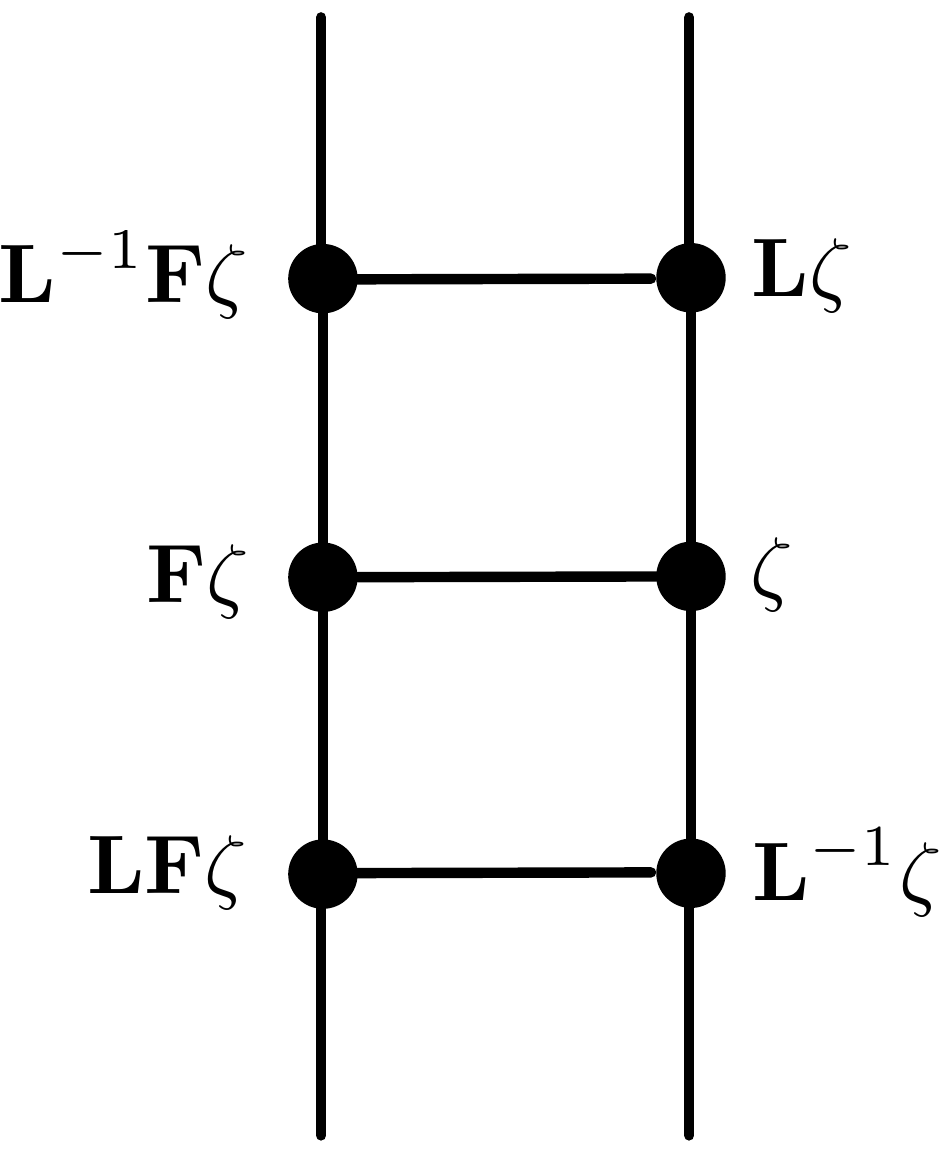}\\(a)\end{tabular}
&
\begin{tabular}{c}\includegraphics[width=0.27\linewidth]{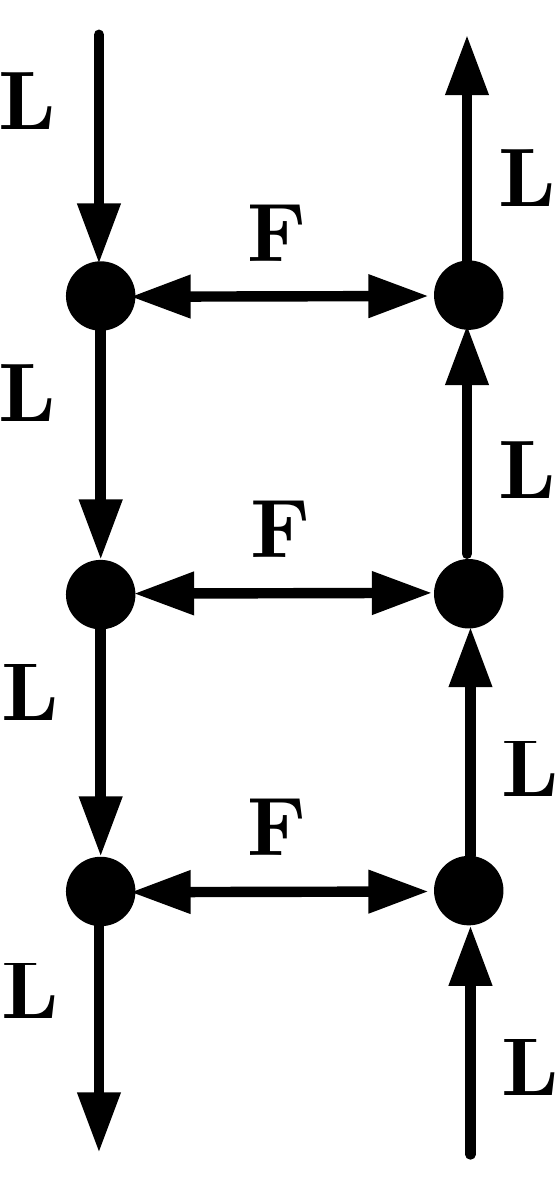}\\(d)\end{tabular}
\end{tabular}
} 
\end{center}
\caption{
This diagram illustrates the possible transitions between states using the Markov transition operator from Equation \ref{trans}.  In \emph{(a)} the relevant states, represented by the nodes, are labeled.  In \emph{(b)} the possible transitions, represented by the arrows, are labeled.    In Section \ref{fixed point}, the net probability flow into and out of the state $\zeta$ is set to 0.
}
\label{transitions}
\end{figure}

\section{Making the distribution of interest a fixed point}\label{fixed point}

In order to make $p\left(\zeta\right)$ a fixed point, we will choose the Markov dynamics $T$ so that on average as many transitions enter as leave state $\zeta$ at equilibrium.  This is {\em not} pairwise detailed balance --- instead we are directly enforcing zero net change in the probability of each state by summing over all allowed transitions into or out of the state.  This constraint is analogous to Kirchhoff's current law, where the total current entering a node is set to 0.  As can be seen from Equation \ref{trans} and the definitions in Section \ref{formalism}, and as is illustrated in Figure \ref{transitions}, 
a state $\zeta$ can only lose probability to the two states $L \zeta$ and $F \zeta$, and gain probability from the two states $L^{-1} \zeta$ and $F^{-1}\zeta$.  Equating the rates of probability inflow and outflow, we find
\begin{align}
p\left( \zeta \right) P_{leap}\left( \zeta \right) +
p\left( \zeta \right) P_{flip}\left( \zeta \right)
&=
p\left( L^{-1}\zeta \right) P_{leap}\left( L^{-1}\zeta \right) +
p\left( F^{-1}\zeta \right) P_{flip}\left( F^{-1}\zeta \right)
\\
&=
p\left( L^{-1}\zeta \right) P_{leap}\left( L^{-1}\zeta \right) +
p\left( \zeta \right) P_{flip}\left( F\zeta \right)
\\
P_{flip}\left( \zeta \right) - P_{flip}\left( F\zeta \right)
&=
\frac{p\left( L^{-1}\zeta \right)}{p\left( \zeta \right) } P_{leap}\left( L^{-1}\zeta \right) - P_{leap}\left( \zeta \right)
\label{balance intermediate}
.
\end{align}

We choose the standard Metropolis-Hastings acceptance rules for $P_{leap}\left( \zeta \right)$,
\begin{align}
P_{leap}\left( \zeta \right) = \min\left( 1, \frac{p\left(L\zeta\right)}{p\left(\zeta\right)} \right)
.
\end{align}
Substituting this in to Equation \ref{balance intermediate}, we find
\begin{align}
P_{flip}\left( \zeta \right) - P_{flip}\left( F\zeta \right)
&=
\frac{p\left( L^{-1}\zeta \right)}{p\left( \zeta \right) } \min\left( 1, \frac{p\left(LL^{-1}\zeta\right)}{p\left(L^{-1}\zeta\right)} \right) - \min\left( 1, \frac{p\left(L\zeta\right)}{p\left(\zeta\right)} \right)
\\
&=
\min\left( 1, \frac{p\left( L^{-1}\zeta \right)}{p\left( \zeta \right) } \right) - \min\left( 1, \frac{p\left(L\zeta\right)}{p\left(\zeta\right)} \right)
\\
&=
\min\left( 1, \frac{p\left( LF\zeta \right)}{p\left( \zeta \right) } \right) - \min\left( 1, \frac{p\left(L\zeta\right)}{p\left(\zeta\right)} \right)
\label{balance}
.
\end{align}
Satisfying Equation \ref{balance} we choose\footnote{To recover standard HMC, instead set $P_{flip}\left( \zeta \right) = 1 - P_{leap}\left( \zeta \right)$.  One can verify by substitution that this satisfies Equation \ref{balance}.}
 the following form for $P_{flip}\left( \zeta \right)$,
\begin{align}
P_{flip}\left( \zeta \right)
&=
\max\left( 0,
\min\left( 1, \frac{p\left( LF\zeta \right)}{p\left( \zeta \right) } \right) - \min\left( 1, \frac{p\left(L\zeta\right)}{p\left(\zeta\right)} \right)
\right)
\label{pflip}
.
\end{align}
Note that $P_{flip}\left( \zeta \right) \leq 1 - P_{leap}\left( \zeta \right)$, where $1 - P_{leap}\left( \zeta \right)$ is the rejection rate, and thus the momentum flip rate, in standard HMC.  Using this form for $P_{flip}\left( \zeta \right)$ will generally reduce the number of momentum flips required.

\begin{figure}
\begin{center}
\parbox[c]{0.6\linewidth}{
\includegraphics[width=1.0\linewidth]{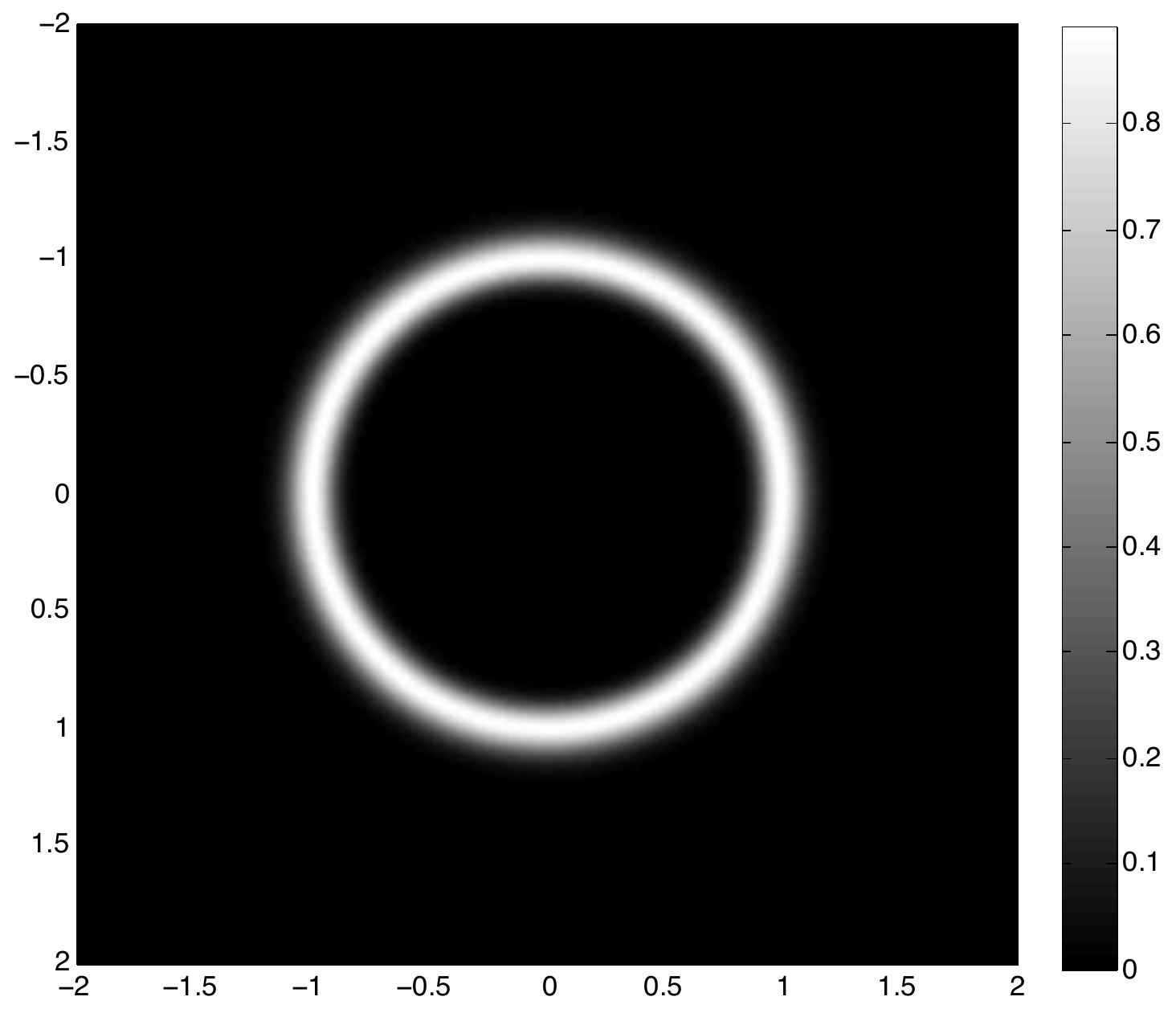}
} 
\end{center}
\caption{
A two dimensional image of the distribution used in Section \ref{ex section}.  Pixel intensity corresponds to the probability density function at that location.
}
\label{distribution}
\end{figure}

\begin{figure}
\begin{center}
\parbox[c]{0.6\linewidth}{
\includegraphics[width=1.0\linewidth]{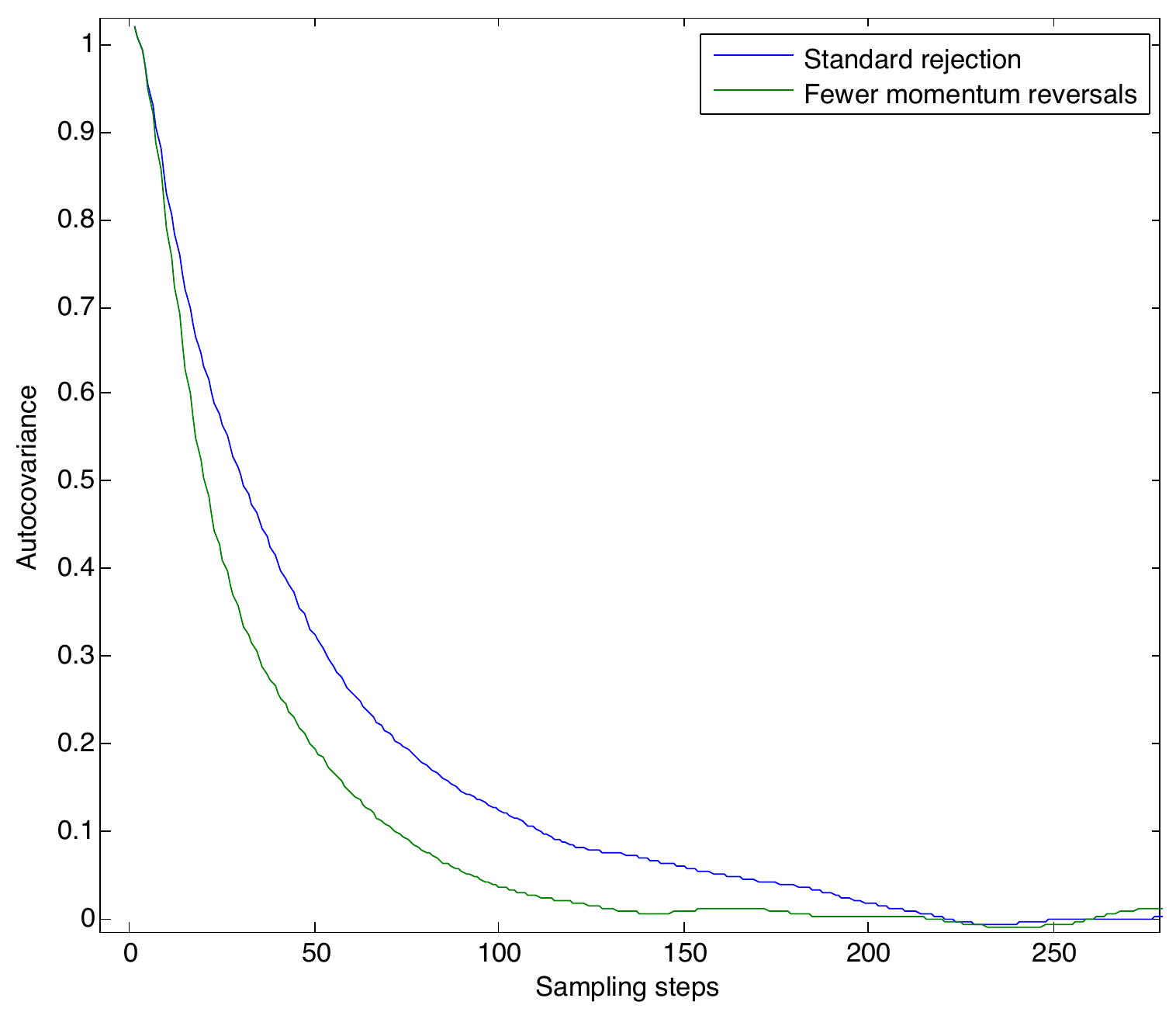}
} 
\end{center}
\caption{
The covariance between samples as a function of the number of intervening sampling steps for HMC with standard rejection and rejection with fewer momentum reversals.  Reducing the number of momentum reversals causes faster mixing, as evidenced by the faster falloff of the autocovariance.
}
\label{autocovariance}
\end{figure}
\section{Example} \label{ex section}

In order to demonstrate the accelerated mixing provided by this technique, samples were drawn from a simple distribution with standard rejection, and with separate rejection and momentum flipping rates as described above.  In both cases, the leapfrog step length $\epsilon$ was set to 0.1, the number of integration steps $n$ was set to 1, and the momentum corruption rate $\beta$ was set so as to corrupt half the momentum per unit stimulation time.  Both samplers were run for $100,000$ sampling steps.  The distribution used was described by the energy function
\begin{align}
E &= 100 \log^2\left( \sqrt{x_1^2 + x_2^2} \right)
.
\end{align}
A 2 dimensional image of this distribution can be seen in Figure \ref{distribution}.  The autocovariance of the returned samples can be seen, as a function of the number of intervening sampling steps, in Figure \ref{autocovariance}.  Sampling using the technique presented here led to more rapid decay of the autocovariance, consistent with faster mixing.



\bibliographystyle{apalike}
\bibliography{note_HMC_reducedflip}

\end{document}